\documentclass[twocolumn,prl]{revtex4}

\usepackage{graphicx}
\usepackage{dcolumn}
\usepackage{amsmath}

\begin{document}
\preprint{HEP/123-qed}

\title[Short Title]{Excitons in Carbon Nanotubes with Broken
Time-Reversal Symmetry}

\author{S. Zaric,$^1$ G. N. Ostojic,$^1$ J. Shaver,$^1$ J.
Kono,$^{1,\dagger}$ O. Portugall,$^2$ P. H. Frings,$^2$ G. L. J.
A. Rikken,$^2$ \\ M. Furis,$^3$ S. A. Crooker,$^3$ X. Wei,$^4$ V.
C. Moore,$^{5}$ R. H. Hauge$^5$} \author{R. E. Smalley$^5$}

\affiliation{$^1$Department of Electrical and Computer
Engineering,
Rice University, Houston, Texas 77005 \\
$^2$Laboratoire National des Champs Magn\'{e}tiques Puls\'{e}s,
31432 Toulouse Cedex 04, France \\
$^3$National High Magnetic Field Laboratory, Los Alamos National
Laboratory, Los Alamos, New Mexico 87545 \\ $^4$National High
Magnetic Field Laboratory, Florida State University,
Tallahassee, Florida 32310 \\ $^5$Department of Chemistry,
Rice University, Houston, Texas 77005}

\date{\today}

\begin{abstract}
Near-infrared magneto-optical spectroscopy of single-walled
carbon nanotubes reveals two absorption peaks with an equal
strength at high magnetic fields ($>$ 55 T).  We show that the
peak separation is determined by the Aharonov-Bohm phase due to
the tube-threading magnetic flux, which breaks the time-reversal
symmetry and lifts the valley degeneracy.
This field-induced symmetry breaking thus overcomes the
Coulomb-induced intervalley mixing which is predicted to make
the lowest exciton state optically inactive (or ``dark'').
\end{abstract}

\pacs{aaa}
\maketitle

A striking prediction for single-walled carbon nanotubes (SWNTs)
in a magnetic field ($B$) parallel to the tube axis is that the
band gap oscillates with period $\phi_0$, the magnetic flux
quantum \cite{AjikiAndo93JPSJ}.  This is due to the appearance
of the Aharonov-Bohm (AB) phase, 2$\pi\phi/\phi_0$ (where $\phi$
is the tube-threading flux), in the circumferential boundary
condition on the electronic wavefunction.  As a result, metallic
tubes become semiconducting even in an infinitesimally small $B$
and semiconducting tubes can become metallic in ultrahigh $B$'s.
Furthermore, the degeneracy between `left-handed' and
`right-handed' electrons having Bloch vectors with opposite
helicity (or the K and K' points in graphene $k$-space) can be
lifted by a $B$, which is expected to appear as
AB-phase-dependent spectral peak splittings and shifts in
interband optical spectra
\cite{AjikiAndo94Physica,RocheetAl00PRB,ShyuetAl03PRB,Ando04JPSJ}.
While low-temperature magneto-transport experiments are strongly
affected by disorder and quantum interference effects
\cite{MagTransport}, magneto-optical studies should be able
to provide more clear-cut evidence of these predictions, with
quantitative information on the splitting rates with $B$.


\begin{figure}
\includegraphics [scale=0.5] {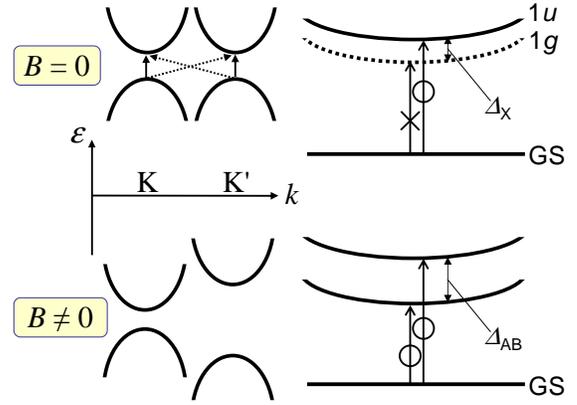}
\caption{The expected $B$-evolution of K-K' intervalley
mixing and splitting in a single-particle picture (left) and an
excitonic picture (right).  1$u$ (1$g$) is the bonding
(anti-bonding) superposition state of the K and K' exciton
states.  The solid (dashed) line represents a bright (dark)
exciton state.  $\Delta_X$: Coulomb-induced splitting;
$\Delta_{AB}$: Aharonov-Bohm-induced splitting.  Both lowest
(singlet) exciton states become bright at high enough magnetic
fields such that $\Delta_{AB} > \Delta_X$.} \label{dark}
\end{figure}

Optical processes are strongly affected by Coulomb
interactions, especially in low-dimensional systems such as
carbon nanotubes.  Interband optical transition energies can
be significantly different from predictions based on simple
band structure models due to quasi-particle corrections and
excitonic shifts, both of which have been shown to be extremely
large in SWNTs
\cite{Ando97JPSJ,SpataruetAl04PRL,ChangetAl04PRL}. There is
growing experimental evidence
\cite{WangetAl05Science,MaultzschetAl05cond-mat} that absorption
and photoluminescence (PL) peaks are excitonic in character,
which is supported by recent theoretical studies
\cite{KaneMele04PRL,Pedersen04Carbon,PerebeinosetAl04PRL,PerebeinosetAl05cond-mat,ZhaoMazumdar04PRL,LouieWONTON}.
One important and unresolved issue is whether optically-inactive,
or ``dark,'' excitons exist.  Such an exciton state is predicted
to exist {\em below} the first optically-active (or ``bright'')
exciton state due to K-K' Coulomb mixing
\cite{ZhaoMazumdar04PRL,PerebeinosetAl05cond-mat,LouieWONTON}
and could explain experimentally-observed low quantum
efficiencies.  A magnetic field can provide insight into this
problem by lifting the K-K' degeneracy in a controllable
manner, especially when it is sufficiently strong that the AB
splitting exceeds the dark-bright energy separation.
Figure \ref{dark} schematically shows the influence of an
applied parallel $B$ on the Coulomb mixing among
the lowest-energy singlet exciton states.  In the presence of
time-reversal symmetry, the lowest two exciton states are the
bonding-like and anti-bonding-like linear combinations of the
K-point and K'-point exciton states, which are expected to split
by an amount $\Delta_X$ determined by the strength of the
electron-hole exchange interaction, the tube diameter, and the
dielectric constant of the surroundings
\cite{PerebeinosetAl05cond-mat}.
The higher (lower) state is predicted to be bright (dark).  When
a symmetry-breaking perturbation (i.e., $B$) is applied, the
K-K' degeneracy is lifted and the importance of the Coulomb
mixing reduces.  At high enough $B$ where $\Delta_{AB} >
\Delta_{X}$, both states become bright and two peaks with an
equal intensity are expected.

Previous magneto-optical studies
\cite{ZaricetAl04Science,ZaricetAl04NL} using high DC magnetic
fields up to 45~T revealed field-induced optical anisotropy and
spectral changes in absorption and PL spectra.  Quantitative
analysis of the PL spectra showed agreement with existing theory
\cite{AjikiAndo93JPSJ,Ando04JPSJ}. However, the expected
absorption splittings in the first-subband region were not
resolved, presumably because the fields used were not large
enough to make the splittings larger than the linewidths. In
addition, no spectral changes were observed in the
second-subband region, where linewidths are even larger.


Here we report on the first observation of clear absorption peak
splittings. We performed interband absorption and PL experiments
in micelle-suspended SWNTs in pulsed $B$ up to 74 T.
At fields above $\sim$55 T, we observe clear splittings of
absorption peaks associated with the lowest-energy interband
transitions in semiconducting SWNTs in addition to the
previously-observed red shifts of PL peaks.
The amounts of splittings and shifts as a function of $B$ are
successfully explained by our model taking into account the
field-dependent band structure as well as the angular
distribution of nanotubes.
These results provide the first clear verification of the
theoretical prediction that the K-K' degeneracy of the lowest
states in semiconducting SWNTs can be lifted by a strong $B$
along the tube axis via the AB phase, which produces two
equally-bright exciton states.


\begin{figure}
\includegraphics [scale=0.63] {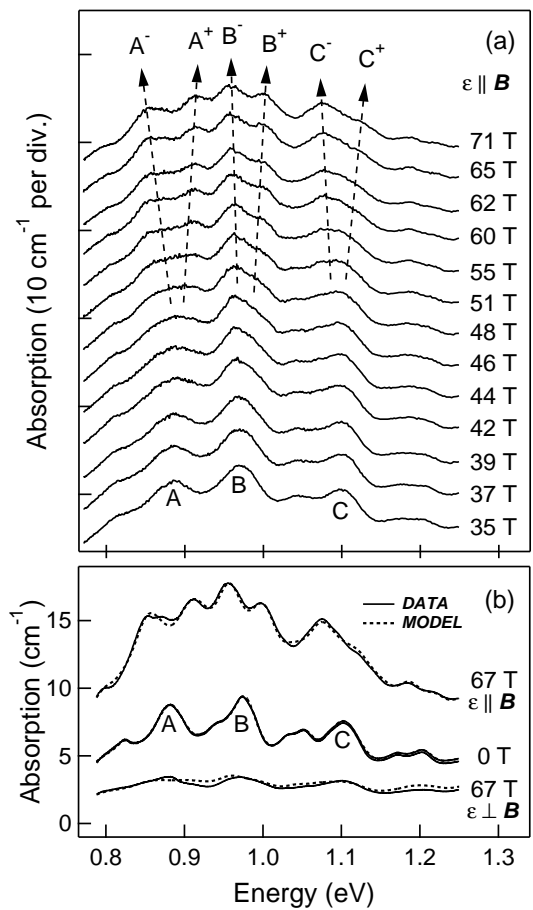}
\caption{Near-bandedge absorption in semiconducting SWNTs in high
magnetic fields for (a) polarization parallel to $B$ (traces are
offset) and (b) both polarizations (no intentional offset).
Each of the three main peaks (labeled A, B, and C) splits into
two at fields above $\sim$55 T.  Dotted lines in (b) are
calculations based on the Aharonov-Bohm effect and magnetic
alignment.} \label{e11} \end{figure}

The samples consisted of SWNTs with diameters 0.6-1.3~nm
suspended in sodium cholate and heavy water.  They were prepared
through homogenization, high power sonication, and
centrifugation, similar to other processes using sodium dodecyl
sulfate \cite{OconnelletAl02Science}.  Such samples are rich in
unbundled nanotubes surrounded by sodium cholate surfactant
molecules, and are thus prevented from interacting with each
other, and show chirality-dependent peaks in absorption and PL
spectra \cite{OconnelletAl02Science,BachiloetAl02Science}.
Lowest-energy interband absorption peaks $E_{11}$ (at 0.8-1.5~eV)
are associated with the transitions between the highest valence
subband and lowest conduction subband in semiconducting SWNTs
with various chiralities while second-subband transitions
$E_{22}$ occur at 1.35-2.25 eV.

We performed magneto-absorption and magneto-PL measurements
in pulsed $B$ fields up to 74~T at room temperature using a
Si charge coupled device (in the visible) and an InGaAs array
detector (in the near-infrared). All measurements were
made in the Voigt geometry, i.e., the light propagation vector
perpendicular to $B$.  Spectra were recorded using a typical
exposure plus readout time of $\sim$1.5~ms.
Measurements were performed in a variety of pulsed magnets. In
Toulouse, magneto-optical absorption measurements were made using
a 75~T magnet (the ARMS magnet \cite{JonesetAl04PhysicaB}).  In
Los Alamos, both absorption and PL studies were performed in a
capacitor-driven 67~T magnet having a 10~ms risetime and an
exponential decay. Absorption studies were also performed using
a prototype insert coil for a planned 100~T Multi-Shot Magnet
\cite{BaconetAl02IEEE}; operating independently, this insert
magnet produced 75~T fields with a 5~ms risetime.

\begin{figure}
\includegraphics [scale=0.65] {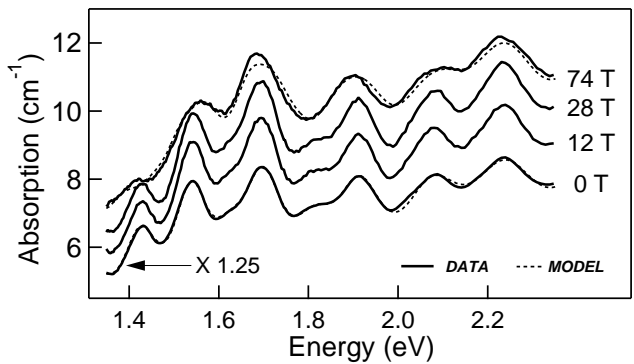}
\caption{Interband magneto-absorption data (solid lines) and
calculations (dashed lines) for light polarized parallel to $B$
in the second-subband range.  Data at non-zero $B$ were taken
during the falling slope of the $B$ pulse.  Each peak becomes
broader with $B$.
} \label{e22}
\end{figure}

Figure \ref{e11}(a) shows $B$-dependent absorption spectra in the
$E_{11}$ region up to 71.4~T. The light was linearly-polarized
along the $B$ direction, i.e.,
$\hat{\varepsilon}\parallel\vec{B}$, where $\hat{\varepsilon}$
is a unit vector in the polarization direction.  The absorption
increases with $B$ in this configuration due to the fact that
the nanotubes align with $B$ \cite{ZaricetAl04Science} and the
$E_{11}$ absorption occurs only for the light polarized parallel
to the tube axis \cite{AjikiAndo94Physica}.
Three main peaks are dominant at zero $B$, which are labelled A,
B, and C.
Here, Peak A ($\sim$0.88~eV) consists primarily of five types of
SWNTs [($n$,$m$) = (9,8), (10,6), (11,6), (12,2), and (11,4)],
Peak B ($\sim$0.98~eV) consists of five types [(10,3), (10,5),
(8,7), (11,1), and (9,5)], and Peak C ($\sim$1.1~eV) consists of
four types [(7,6), (8,4), (9,2), and (9,4)].  At fields above
$\sim$55~T, each of these peaks splits into two clearly resolved
peaks (e.g., Peaks A$^+$ and A$^-$). Figure \ref{e11}(b) shows 0
and 67~T absorption spectra (solid lines) for parallel and
perpendicular polarizations together with theoretical curves
(dashed lines) obtained through our model (described later).  No
traces are intentionally offset, and it is seen that absorption
increases (decreases) with $B$ for parallel (perpendicular) to
the polarization, and the splittings are visible only in the
parallel case.


Figure \ref{e22} shows the $B$-dependence of
$\hat{\varepsilon}\parallel\vec{B}$ data (solid lines) and
calculations (dashed lines) in the second-subband ($E_{22}$)
region. Again, there are two main $B$-induced effects. First,
absorption increases with increasing $B$ due to magnetic
alignment. Second, absorption peaks show broadening at high
fields,
but no clear splitting is observed even at the highest $B$.  This
is due to the much broader linewidths ($\sim$100 meV) of $E_{22}$
transitions compared with $E_{11}$ transitions (20-30 meV). The
large linewidths of $E_{22}$ transitions are consistent with
short intraband relaxation times \cite{OstojicetAl04PRL}.

\begin{figure}
\includegraphics [scale=0.68] {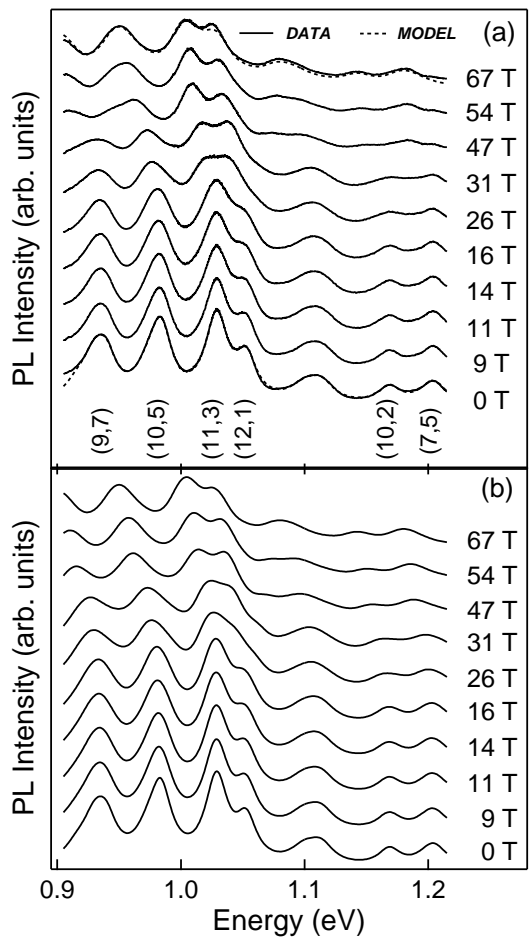}
\caption{(a) Photoluminescence spectra (solid lines) at various
$B$ with 792.5 nm excitation. Data at non-zero $B$
were taken during the falling slope of the $B$ pulse.
The dashed lines are simulations.  (b) Theoretical
simulations of the magneto-PL data shown in (a).} \label{pl}
\end{figure}

The $B$ evolution of PL spectra up to 67~T taken with 792.5 nm
excitation is shown in Fig.~\ref{pl}(a).  Peaks are labeled
by the corresponding dominant chiralities ($n$,$m$).  Here a
highly nonlinear $B$ dependence is observed, i.e., drastic
spectral changes occur only at high enough magnetic fields.  At
fields from 0~T to 16~T, each peak slightly broadens with
increasing $B$, but their peak positions remain the same. With
further increasing $B$, however, all peaks start shifting to
lower energies rather abruptly and continue to red shift with
$B$ up to the highest field (67~T).  Furthermore, concomitantly,
their linewidths decrease. All the traces in Fig.~\ref{pl}(b) and
the dotted lines in \ref{pl}(a) are theoretical curves, which
successfully simulate the corresponding experimental curves in
Fig.~\ref{pl}(a).  As detailed below, these simulations were
obtained by taking into account the AB-effect-induced band
structure modifications together with magnetic alignment.

For a nanotube in a parallel $B$, the amount of AB-induced K-K'
splitting $\Delta_{AB}$ is expected to be proportional to $B$,
i.e., $\Delta_{AB} \equiv vB$, when $\phi/\phi_0 \ll 1$.
Here the rate of splitting $v$ is chirality-dependent and, for a
1-nm-diameter nanotube, $v$ is expected to be
$\sim$1~meV/T \cite{AjikiAndo93JPSJ,Ando04JPSJ}.
However, experimentally, we are measuring peak splittings and
shifts for an {\it ensemble} of nanotubes with different angles
($\theta$) with respect to the $B$ direction. Using
Maxwell-Boltzmann statistics under equilibrium conditions, the
probability $P(\theta)d\theta$ of finding a nanotube in an
angular range of $(\theta,\theta+d\theta)$ can be written in
spherical coordinates as
\begin{eqnarray}
P_{u}(\theta)d\theta = {\exp(-u^2 \sin^2\theta)\sin\theta d\theta
\over \int^{\pi/2}_{0}\exp(-u^2 \sin^2\theta)\sin\theta d\theta}
\ , \\ u \equiv [B^2 N (\chi_{\parallel} - \chi_{\perp})/k_B
T]^{1/2} \ , \label{prob}\end{eqnarray}
where $N$ is the number of moles of carbon atoms in the nanotube,
and $\chi_{\parallel}$ and $\chi_{\perp}$ are parallel and
perpendicular diamagnetic susceptibilities per mole of carbon
atoms.
Using this distribution function, we modeled the PL data
measured with two excitation wavelengths (792.5~nm and 750~nm) in
the same way as in our previous work \cite{ZaricetAl04Science}.
The 0~T spectrum was fitted using Lorentzians that represent the
various chiralities present in the sample.  The 67~T PL spectrum
was then simulated by varying the two parameters $u$ and $v$ for
each Lorentzian with the linewidth determined at 0~T.
For any given $\theta$, $v$ was multiplied by $\cos\theta$ to
account only for the tube-threading component of the magnetic
flux. The relative intensities of the split peaks were calculated
through the Boltzmann factor $\exp(-\Delta_{AB}\cos\theta/k_B
T)$ with $T$ = 300~K, where $k_B$ is the Boltzmann constant.
Averaging over $\theta$ was performed with a weight equal to
$P_u(\theta)$.  The average splitting rate obtained in this way
was $v$ = 0.9 meV/T.  Data at intermediate fields (0 $< B <$ 67
T) were then simulated without any adjustable parameters, as
shown in Fig.~\ref{pl}(b).  Here, for each peak at each $B$, the
value of $u$ was calculated from the 67~T value through the
$u$-$B$ proportionality, i.e., Eq.~(2), and there is good
overall agreement with the data in Fig.~\ref{pl}(a).
The obtained $v$ agrees with our low-$B$ ($<$45~T) results
\cite{ZaricetAl04Science}, and
the successful simulation of the PL data demonstrates that,
although the nanotubes are not exactly under equilibrium
conditions due to the use of pulsed $B$, Eq.~(1) still
adequately describes the angular distribution.


In PL, the relative intensities of the split peaks
are affected by the population difference, and the higher energy
peak becomes less and less visible as $B$ (and thus
$\Delta_{AB}$) increases.
Split {\em absorption} peaks, on the other hand, should have the
same oscillator strength.
To test this, we modeled the absorption data in an essentially
the same manner as for the PL data but {\em without a Boltzmann
factor}.  We took into account the effect of the polarizer used
in the experiment through a factor of $\cos^2\alpha$, where
$\alpha$ is the angle between the tube axis and the polarizer
axis.  The model works well for the 67~T absorption data in the
$\hat{\varepsilon}\parallel\vec{B}$ case [see Fig.~\ref{e11}(b)]
with $u_{avg}$ = 1.9 and $v$ = 0.7~meV/T \cite{align}.  Using
the same values of $u$ and $v$, the 67~T
$\hat{\varepsilon}\perp\vec{B}$ absorption data was also
reproduced successfully [see Fig.~\ref{e11}(b)]. The same model
was then applied to the absorption in the second subband region
and, as shown in Fig.~\ref{e22}, the broadenings observed were
well explained.
These results confirm that there are two equally-bright exciton
states at high $B$ ($>$ 55~T).

In summary, we have performed magneto-optical experiments on
semiconducting SWNTs in pulsed magnetic fields up to 74~T.
Clear splittings of absorption peaks in the first-subband region
as well as significant redshifts of PL peaks were observed as a
consequence of the Aharonov-Bohm phase changing the band
structure.
The observation of clear absorption peak splittings at high
magnetic fields not only provides direct evidence of the
expected K-K' splitting due to the $B$-induced symmetry breaking
but also sheds some light on the issue of ``dark'' excitons.
Recent theories predict that one of the lowest two excitons is
dark.  Our data demonstrates that both lowest-energy excitons
are bright when a high enough magnetic field is applied
along the tube axis.

We thank V.~I.~Klimov for the use of his InGaAs
detector for measurements in Los Alamos.  We also thank
T.~Ando, E.~K.~Chang, S.~G.~Louie, E.~J.~Mele, and
V.~Perebeinos for useful discussions on dark excitons.  This
work was supported in part by the Robert A.~Welch Foundation
(through Grant No.~C-1509) and the National Science Foundation
(through Grant Nos.~DMR-0134058, DMR-0325474, and INT-0437342).

\bigskip

\noindent$^{\dagger}$To whom correspondence should be addressed.
Electronic address: kono@rice.edu.


\begin{thebibliography}{25}
\expandafter\ifx\csname natexlab\endcsname\relax\def\natexlab#1{#1}\fi
\expandafter\ifx\csname bibnamefont\endcsname\relax
  \def\bibnamefont#1{#1}\fi
\expandafter\ifx\csname bibfnamefont\endcsname\relax
  \def\bibfnamefont#1{#1}\fi
\expandafter\ifx\csname citenamefont\endcsname\relax
  \def\citenamefont#1{#1}\fi
\expandafter\ifx\csname url\endcsname\relax
  \def\url#1{\texttt{#1}}\fi
\expandafter\ifx\csname urlprefix\endcsname\relax\def\urlprefix{URL }\fi
\providecommand{\bibinfo}[2]{#2}
\providecommand{\eprint}[2][]{\url{#2}}

\bibitem[{\citenamefont{Ajiki and Ando}(1993)}]{AjikiAndo93JPSJ}
\bibinfo{author}{\bibfnamefont{H.}~\bibnamefont{Ajiki}} \bibnamefont{and}
  \bibinfo{author}{\bibfnamefont{T.}~\bibnamefont{Ando}}, \bibinfo{journal}{J.
  Phys. Soc. Jpn.} \textbf{\bibinfo{volume}{62}}, \bibinfo{pages}{1255}
  (\bibinfo{year}{1993}).

\bibitem[{\citenamefont{Ajiki and Ando}(1994)}]{AjikiAndo94Physica}
\bibinfo{author}{\bibfnamefont{H.}~\bibnamefont{Ajiki}} \bibnamefont{and}
  \bibinfo{author}{\bibfnamefont{T.}~\bibnamefont{Ando}},
  \bibinfo{journal}{Physica B} \textbf{\bibinfo{volume}{201}},
  \bibinfo{pages}{349} (\bibinfo{year}{1994}).

\bibitem[{\citenamefont{Roche et~al.}(2000)\citenamefont{Roche, Dresselhaus,
  Dresselhaus, and Saito}}]{RocheetAl00PRB}
\bibinfo{author}{\bibfnamefont{S.}~\bibnamefont{Roche}} {\it et
al}.,
\bibinfo{journal}{Phys. Rev. B} \textbf{\bibinfo{volume}{62}},
\bibinfo{pages}{16092} (\bibinfo{year}{2000}).

\bibitem[{\citenamefont{Shyu et~al.}(2003)\citenamefont{Shyu, Chang, Chen,
  Chiu, and Lin}}]{ShyuetAl03PRB}
\bibinfo{author}{\bibfnamefont{F.~L.} \bibnamefont{Shyu}} {\it et
al}.,
  \bibinfo{journal}{Phys. Rev. B} \textbf{\bibinfo{volume}{67}},
  \bibinfo{pages}{045405} (\bibinfo{year}{2003}).

\bibitem[{\citenamefont{Ando}(2004)}]{Ando04JPSJ}
\bibinfo{author}{\bibfnamefont{T.}~\bibnamefont{Ando}}, \bibinfo{journal}{J.
  Phys. Soc. Jpn.} \textbf{\bibinfo{volume}{73}}, \bibinfo{pages}{3351}
  (\bibinfo{year}{2004}).


\bibitem{MagTransport}
See, e.g., B. Stojetz {\it et al}.,
Phys. Rev. Lett. {\bf 94}, 186802 (2005), G. Fedorov {\it et
al}.,
Phys. Rev. Lett. {\bf 94},
066801 (2005), and references cited therein.

\bibitem[{\citenamefont{Ando}(1997)}]{Ando97JPSJ}
\bibinfo{author}{\bibfnamefont{T.}~\bibnamefont{Ando}}, \bibinfo{journal}{J.
  Phys. Soc. Jpn.} \textbf{\bibinfo{volume}{66}}, \bibinfo{pages}{1066}
  (\bibinfo{year}{1997}).


\bibitem[{\citenamefont{Spataru et~al.}(2004)\citenamefont{Spataru,
  Ismail-Beigi, Benedict, and Louie}}]{SpataruetAl04PRL}
\bibinfo{author}{\bibfnamefont{C.~D.} \bibnamefont{Spataru}},
  \bibinfo{author}{\bibfnamefont{S.}~\bibnamefont{Ismail-Beigi}},
  \bibinfo{author}{\bibfnamefont{L.~X.} \bibnamefont{Benedict}},
  \bibnamefont{and} \bibinfo{author}{\bibfnamefont{S.~G.} \bibnamefont{Louie}},
  \bibinfo{journal}{Phys. Rev. Lett.} \textbf{\bibinfo{volume}{92}},
  \bibinfo{pages}{077402} (\bibinfo{year}{2004}).

\bibitem[{\citenamefont{Chang et~al.}(2004)\citenamefont{Chang, Bussi, Ruini,
  and Molinari}}]{ChangetAl04PRL}
\bibinfo{author}{\bibfnamefont{E.}~\bibnamefont{Chang}} {\it et
al}.,
  \bibinfo{journal}{Phys. Rev. Lett.} \textbf{\bibinfo{volume}{92}},
  \bibinfo{pages}{196401} (\bibinfo{year}{2004}).

\bibitem[{\citenamefont{Wang et~al.}(2005)\citenamefont{Wang, Dukovic, Brus,
  and Heinz}}]{WangetAl05Science}
\bibinfo{author}{\bibfnamefont{F.}~\bibnamefont{Wang}} {\it et
al}.,
  \bibinfo{journal}{Science} \textbf{\bibinfo{volume}{308}},
  \bibinfo{pages}{838} (\bibinfo{year}{2005}).

\bibitem[{\citenamefont{Maultzsch et~al.}(2005)\citenamefont{Maultzsch,
  Pomraenke, Reich, Chang, Prezzi, Ruini, Molinari, Strano, Thomsen, and
  Lienau}}]{MaultzschetAl05cond-mat}
\bibinfo{author}{\bibfnamefont{J.}~\bibnamefont{Maultzsch}} {\it et
al}.,
  \bibinfo{journal}{cond-mat/0505150}.



\bibitem[{\citenamefont{Kane and Mele}(2004)}]{KaneMele04PRL}
\bibinfo{author}{\bibfnamefont{C.~L.} \bibnamefont{Kane}} \bibnamefont{and}
  \bibinfo{author}{\bibfnamefont{E.~J.} \bibnamefont{Mele}},
  Phys. Rev. Lett. {\bf 90}, 207401 (2003);
  {\it ibid}. \textbf{\bibinfo{volume}{93}},
  \bibinfo{pages}{197402} (\bibinfo{year}{2004}).


\bibitem{Pedersen04Carbon}
T. G. Pedersen, Carbon {\bf 42}, 1007 (2004).

\bibitem[{\citenamefont{Perebeinos et~al.}(2004)\citenamefont{Perebeinos,
  Tersoff, and Avouris}}]{PerebeinosetAl04PRL}
\bibinfo{author}{\bibfnamefont{V.}~\bibnamefont{Perebeinos}},
  \bibinfo{author}{\bibfnamefont{J.}~\bibnamefont{Tersoff}}, \bibnamefont{and}
  \bibinfo{author}{\bibfnamefont{P.}~\bibnamefont{Avouris}},
  \bibinfo{journal}{Phys. Rev. Lett.} \textbf{\bibinfo{volume}{92}},
  \bibinfo{pages}{257402} (\bibinfo{year}{2004}).


\bibitem[{\citenamefont{Zhao and Mazumdar}(2004)}]{ZhaoMazumdar04PRL}
\bibinfo{author}{\bibfnamefont{H.}~\bibnamefont{Zhao}} \bibnamefont{and}
  \bibinfo{author}{\bibfnamefont{S.}~\bibnamefont{Mazumdar}},
  \bibinfo{journal}{Phys. Rev. Lett.} \textbf{\bibinfo{volume}{93}},
  \bibinfo{pages}{157402} (\bibinfo{year}{2004}).


\bibitem[{\citenamefont{Perebeinos et~al.}(2005)\citenamefont{Perebeinos,
  Tersoff, and Avouris}}]{PerebeinosetAl05cond-mat}
\bibinfo{author}{\bibfnamefont{V.}~\bibnamefont{Perebeinos}} {\it et
al}.,
  \bibinfo{journal}{cond-mat/0506775}.

\bibitem{LouieWONTON}
C. D. Spataru, S. Ismail-Beigi, R. B. Capaz, and S. G. Louie,
cond-mat/0507067.

\bibitem[{\citenamefont{Zaric et~al.}(2004{\natexlab{a}})\citenamefont{Zaric,
  Ostojic, Kono, Shaver, Moore, Strano, Hauge, Smalley, and
  Wei}}]{ZaricetAl04Science}
\bibinfo{author}{\bibfnamefont{S.}~\bibnamefont{Zaric}} {\it et
al}.,
  \bibinfo{journal}{Science} \textbf{\bibinfo{volume}{304}},
  \bibinfo{pages}{1129} (\bibinfo{year}{2004}{\natexlab{a}}).

\bibitem[{\citenamefont{Zaric et~al.}(2004{\natexlab{b}})\citenamefont{Zaric,
  Ostojic, Kono, Shaver, Moore, Strano, Hauge, Smalley, and
  Wei}}]{ZaricetAl04NL}
\bibinfo{author}{\bibfnamefont{S.}~\bibnamefont{Zaric}} {\it et
al}.,
  \bibinfo{journal}{Nano Lett.} \textbf{\bibinfo{volume}{4}},
  \bibinfo{pages}{2219} (\bibinfo{year}{2004}{\natexlab{b}}).


\bibitem[{\citenamefont{O'Connell et~al.}(2002)\citenamefont{O'Connell,
  Bachilo, Huffman, Moore, Strano, Haroz, Rialon, Boul, Noon, Kittrell
  et~al.}}]{OconnelletAl02Science}
\bibinfo{author}{\bibfnamefont{M.~J.} \bibnamefont{O'Connell}} {\it et
al}.,
\bibinfo{journal}{Science}
  \textbf{\bibinfo{volume}{297}}, \bibinfo{pages}{593} (\bibinfo{year}{2002}).

\bibitem[{\citenamefont{Bachilo et~al.}(2002)\citenamefont{Bachilo, Strano,
  Kittrell, Hauge, Smalley, and Weisman}}]{BachiloetAl02Science}
\bibinfo{author}{\bibfnamefont{S.~M.} \bibnamefont{Bachilo}} {\it et
al}.,
\bibinfo{journal}{Science}
  \textbf{\bibinfo{volume}{298}}, \bibinfo{pages}{2361} (\bibinfo{year}{2002}).


\bibitem[{\citenamefont{Jones et~al.}(2004)\citenamefont{Jones, Frings, von
  Ortenberg, Lagutin, Bockstal, Portugall, and Herlach}}]{JonesetAl04PhysicaB}
\bibinfo{author}{\bibfnamefont{H.}~\bibnamefont{Jones}} {\it et
al}.,
  \bibinfo{journal}{Physica B} \textbf{\bibinfo{volume}{346-347}},
  \bibinfo{pages}{553} (\bibinfo{year}{2004}).

\bibitem[{\citenamefont{Bacon et~al.}(2002)\citenamefont{Bacon, Ammerman, Coe,
  Ellis, Lesch, Sims, Schillig, and Swenson}}]{BaconetAl02IEEE}
\bibinfo{author}{\bibfnamefont{J.~L.} \bibnamefont{Bacon}} {\it et
al}.,
\bibinfo{journal}{IEEE Trans. Appl. Supercond.}
  \textbf{\bibinfo{volume}{12}}, \bibinfo{pages}{695} (\bibinfo{year}{2002}).

\bibitem[{\citenamefont{Ostojic et~al.}(2004)\citenamefont{Ostojic, Zaric,
  Kono, Strano, Moore, Hauge, and Smalley}}]{OstojicetAl04PRL}
\bibinfo{author}{\bibfnamefont{G.~N.} \bibnamefont{Ostojic}} {\it et
al}.,
  \bibinfo{journal}{Phys. Rev. Lett.} \textbf{\bibinfo{volume}{92}},
  \bibinfo{pages}{117402} (\bibinfo{year}{2004});
%
  {\it ibid}. \textbf{\bibinfo{volume}{94}},
  \bibinfo{pages}{097401} (\bibinfo{year}{2005}).






\bibitem{align}
The value for $v$ deduced from 67~T absorption is
smaller than that obtained from 67~T PL.
This is most likely due to the chirality dependence of
$v$.  Absorption peaks are dominated by chiralities that are
relatively abundant in the sample while PL probes only a limited
number of chiralities (using only two excitation wavelengths in
the current study). Thus, more PL data is needed to elucidate
this point.




\end{thebibliography}

\end{document}